\documentclass[12pt,preprint]{aastex}

\shorttitle{On the Nature of the EUV Late Phase of Solar Flares}
\shortauthors{Li et al.}

\begin{document}

\title{On the Nature of the Extreme-Ultraviolet Late Phase \\of Solar Flares}
\author{Y. Li$^{1,2}$, M. D. Ding$^{1,2}$, Y. Guo$^{1,2}$, Y. Dai$^{1,2}$}

\affil{$^1$School of Astronomy and Space Science, Nanjing University, Nanjing 210093, China}
\affil{$^2$Key Laboratory for Modern Astronomy and Astrophysics (Nanjing University), Ministry of Education, Nanjing 210093, China}
\email{yingli@nju.edu.cn}

\begin{abstract}
The extreme-ultraviolet (EUV) late phase of solar flares is a second peak of warm coronal emissions (e.g., Fe {\sc xvi}) for many minutes to a few hours after the {\em GOES} soft X-ray peak. It was first observed by the EUV Variability Experiment (EVE) on board the {\em Solar Dynamics Observatory} ({\em SDO}). The late phase emission originates from a second set of longer loops (late phase loops) that are higher than the main flaring loops. It is suggested as being caused by either additional heating or long-lasting cooling. In this paper, we study the role of long-lasting cooling and additional heating in producing the EUV late phase using the ``enthalpy-based thermal evolution of loops'' (EBTEL) model. We find that a long cooling process in late phase loops can well explain the presence of the EUV late phase emission, but we cannot exclude the possibility of additional heating in the decay phase. Moreover, we provide two preliminary methods based on the UV and EUV emissions from the Atmospheric Imaging Assembly (AIA) on board {\em SDO} to determine whether an additional heating plays some role or not in the late phase emission. Using nonlinear force-free field modeling, we study the magnetic configuration of the EUV late phase. It is found that the late phase can be generated either in hot spine field lines associated with a magnetic null point or in large-scale magnetic loops of multipolar magnetic fields. In this paper, we also discuss why the EUV late phase is usually observed in warm coronal emissions and why the majority of flares do not exhibit an EUV late phase.
\end{abstract}


\section{Introduction}

Solar flares represent the most dynamic events in the solar atmosphere. They involve a variety of physical processes such as  energy release, plasma heating, particle acceleration, mass flows, and waves. In the standard solar flare model, magnetic reconnection occurs in the corona, releasing a large amount of energy \citep{tsun96,prie02}. The released energy is transported downward by non-thermal particles and/or thermal conduction, and heats the lower atmosphere \citep{acto82,mill09}. Chromospheric material is heated and driven into the corona \citep{anto83,bros13}, which fills the flare loops that then brighten up in soft X-ray and extreme-ultraviolet (EUV) wavelength bands. 

In general, the EUV emission of solar flares exhibits one main peak several minutes after the {\em GOES} soft X-ray peak. Recently, however, \cite{wood11} discovered a second peak using the EUV irradiance observations from the EUV Variability Experiment (EVE; \citealt{wood12}) on board the {\em Solar Dynamics Observatory} ({\em SDO}; \citealt{pesn12}). This secondary peak, referred to as the EUV late phase, lags behind the first peak by many minutes or even a few hours. \cite{wood11} presented four criteria for this EUV late phase of solar flares: (1) a second peak in the warm emissions (e.g., Fe {\sc xv} and Fe {\sc xvi}) after the {\em GOES} soft X-ray peak, (2) no significant enhancements in hot emissions (e.g., Fe {\sc xx}) during the second peak, (3) association with an eruptive event seen in EUV images, and (4) existence of a second set of higher and longer loops. The EUV late phase of flares mainly shows up in warm coronal emissions, most often in the EVE Fe {\sc xvi} 33.5 nm ($\sim$3 MK) and Fe {\sc xv} 28.4 nm ($\sim$2 MK) lines, and sometimes in the EVE Fe {\sc xviii} 9.4 nm ($\sim$6 MK) line. This emission is usually accompanied by an eruption and resides in a second set of longer loops higher than the main flaring loops. Note that EUV late phase flares are distinct from long-duration flares, and that not all flares have an EUV late phase \citep{wood11,wood14}. 

It has been found that the EUV late phase emission comes from higher and larger loop systems (called late phase loops hereafter) rather than the main flaring loops using high-spatial-resolution EUV images from the Atmospheric Imaging Assembly (AIA; \citealt{leme12}) on board {\em SDO} \citep{wood11, hock12}. A number of case studies proposed two physical explanations for EUV late phase emission: one is an additional heating in the late phase loops \citep{wood11, hock12, daiy13, sunx13}, and the other is long-lasting cooling of late phase loops \citep{liuk13,sunx13}. In the first explanation, the EUV late phase is explained as the natural result of a separate energy release and heating event with a lower heating rate than the flare main phase; while in the second explanation, the EUV late phase is due to different cooling times in flare loops with different lengths. The late phase loops are much longer in length and also in cooling time \citep{curd04} than the main flaring loops. The superposition of emissions of these two kinds of loop results in a hump in warm coronal emissions in the late  phase.

Which mechanism, additional heating or long-lasting cooling, or both, plays the key role in the EUV late phase? How much do they contribute to the late phase? Any mechanism needs to answer two important questions related to the EUV late phase. First, why is the EUV late phase observed in warm coronal emissions (e.g., Fe {\sc xvi}) but not in hot coronal emissions (e.g., Fe {\sc xxiii})? Second, what is the typical magnetic field configuration involved in the EUV late phase emission? Up to now, these questions have not been fully resolved.

Previous studies on the EUV late phase basically started with observations from EVE and AIA. In fact, there exist limitations on studying the intrinsic mechanism for the late phase from observations. For example, EVE has no spatial resolution. Even for AIA, superposition of different loops along the line of sight and saturation in some channels for large flares (most of the EUV late phase flares are relatively large) make it difficult to infer the exact emission from each loop. Considering all of these, we study the EUV late phase of solar flares through numerical experiments using the ``enthalpy-based thermal evolution of loops'' model (EBTEL; \citealt{klim08,carg12}) and nonlinear force-free field modeling (NLFFF; \citealt{2000Wheatland}; \citealt{2004Wiegelmann}), which can shed light on the nature of the EUV late phase when combined with {\em SDO} observations. This paper is organized as follows: in Section \ref{overview}, we describe the observational properties of the EUV late phase of flares; then, in Section \ref{experiments} we show our EBTEL experiments based on observations; in Section \ref{nature}, we clarify the nature of different aspects of the EUV late phase; finally, we present a summary and some discussions in Section \ref{discussion}.

\section{Observational Properties of the EUV Late Phase}
\label{overview}

\cite{wood11} presented a list of EUV late phase flares brighter than C2 during May 2010--March 2011. A number of papers have been published each of which focuses on a particular event of the EUV late phase. Here we summarize the observational properties of five EUV late phase flares, which have been carefully analyzed by different authors, in Table \ref{table}. 

\begin{table}[htb]
\begin{center}
\tiny
\caption{Observational Properties of Five EUV Late Phase Flares}
\label{table}
\begin{tabular}{cccccccccc}
\tableline
\tableline
 Date & NOAA & \multicolumn{2}{c}{{\em GOES} Soft X-ray} & Late Phase & Peak & 
 \multicolumn{2}{c}{Loop Length$^{\rm c}$ (Mm) } & Explanation$^{\rm d}$ & References \\
\cline{3-4} \cline{7-8}
     &  AR   & Flare Class & Peak Time & Delay Time$^{\rm a}$  & Ratio$^{\rm b}$  
& Main Flaring & Late Phase &   \\
     &         &                    & (UT)             & (min)  & & Loop & Loop &   \\
\tableline
 5-May-2010 & 11069 & C8.8 & 11:52 & 78 &  0.6  & 20 & 70 & heating & \cite{hock12} \\
16-Oct-2010 & 11112 & M2.9 & 19:12 & 61 & 0.15 & 28 & 87 & cooling & \cite{liuk13} \\
18-Feb-2011 & 11158 & M1.4 & 13:03 & 42 & 0.36 & 22 & 72 & cooling & \cite{liuk13} \\
  6-Sep-2011 & 11283 & X2.1  & 22:20 & 83 & 0.09 & 35 &100& heating & \cite{daiy13} \\
15-Nov-2011 & 11346 & M1.9 & 12:43 & 75 & 0.8  & 30 & 150& cooling & \cite{sunx13} \\
 & & & & & &  & & (+heating) & \\
\tableline
\end{tabular}
\end{center}
\small
$^{\rm a}$ The delay time is the time interval from the {\em GOES} soft X-ray peak to the second peak of the EVE Fe {\sc xvi} 33.5 nm emission.\\
$^{\rm b}$ It is the ratio of the second peak to the first (main) peak in the EVE Fe {\sc xvi} 33.5 nm emission.\\
$^{\rm c}$ The loop length is estimated from AIA images.\\
$^{\rm d}$ ``Heating'' represents an additional heating, and ``cooling'' represents long time cooling.
\end{table}

From Table \ref{table}, it is seen that these EUV late phase flares are relatively large (larger than C8), and that the late phase is delayed with respect to the main flare peak by about 1 hour. The relative peak emission of the late phase varies greatly in magnitude. In some cases, it can be very small ($\sim$0.1 times the main peak emission) and in other cases, it can be fairly large (0.8 times the main peak emission). Late phase loops are about 2--4 times longer than main flaring loops. In these events, some authors explained the EUV late phase as being caused by additional heating, and some proposed that it is the result of a long cooling process. To address the origin of the EUV late phase emission, some authors have also used the EBTEL model to simulate their target events and found results in support of an additional heating scenario \citep{hock12,sunx13}. 

Besides emission, the magnetic configuration of the EUV late phase has also been studied in previous papers on the late phase. \cite{wood11} reported that the magnetic field configuration of the active region (AR) seems to be important for the EUV late phase, and that a topologically distinct set of loops is involved during the late phase flare. \cite{hock12} proposed a physical model---a classic quadrupolar configuration based on breakout reconnection, which is similar to the coronal mass ejection (CME) initiation model in \cite{lync08}. \cite{liuk13} provided an asymmetric quadruple magnetic topology model, in which a sigmoid core produces the main phase, while rising AR fields interact and reconnect with the overlying large loop arcade, and the large arcade cools down slowly and produces the observed EUV late phase. \cite{daiy13} proposed a three-stage magnetic reconnection scenario under a multipolar magnetic topology, and ascribed the EUV late phase as a product of the third stage. Quite recently, \cite{sunx13} reported a unique fan-spine magnetic topology, and showed that the EUV late phase emission comes from large overlying post-reconnection loops, which are naturally formed in such a fan-spine structure. According to these studies, it is seen that a multipolar magnetic topology is very important for the appearance of an EUV late phase.

\section{Numerical Experiments}
\label{experiments}

Previous studies on the EUV late phase of solar flares were mainly based on {\em SDO} observations. Some of them also involved simulations (e.g., \citealt{hock12,sunx13}) but they just aimed to explain a specific late phase event. Here we conduct numerical experiments using the EBTEL model to study the EUV late phase in more detail, concentrating on, in particular, the role of additional heating and long-lasting cooling in the EUV late phase.

\subsection{The EBTEL Model}

The EBTEL model is a zero-dimensional (0D) and highly efficient model, which describes the evolution of mean properties (e.g., temperature, density, pressure, and enthalpy flow velocity) of coronal plasma. It solves these time-dependent equations for density and pressure:
\begin{equation}
 \frac{dn}{dt}=-\frac{2\,c_2}{5\,k_B\,T}\,\left\lbrack\frac{F_c}{L}+c_1\,n^2\,\Lambda(T)\right\rbrack
 \label{eq:n}
\end{equation}
and  
\begin{equation}
 \frac{dP}{dt}=\frac{2}{3}\,\left\lbrack Q(t)-(1+c_1)\,n^2\,\Lambda(T)\right\rbrack,
\label{eq:p}
\end{equation}
where $n$, $P$, and $T$ are mean density, pressure, and temperature of the loop, respectively, $P=2\,n\,k_B\,T$, $k_B$ is the Boltzmann constant, $c_2=0.9$ is a typical ratio of mean temperature to apex temperature of a coronal loop, $L$ is the loop half-length (EBTEL assumes a symmetric loop), $Q(t)$ is the volumetric heating rate, $\Lambda(T)$ is the radiative loss function, $F_c=-\frac{2}{7}\,\kappa_0\,(T/c_2)^{7/2}/L$ is the thermal conductive flux, and $c_1$ is the ratio of radiative loss through the transition region to coronal radiative loss. Note that here we use the new version of the EBTEL model described in \cite{carg12}, in which $c_1$ is consistently calculated rather than being held constant as in the earlier version of \cite{klim08}. In addition, we only consider thermal heating $Q(t)$ and ignore non-thermal beam heating. Doing so is reasonable to study the late phase emission of solar flares for the following reasons. In the additional heating scenario, the non-thermal heating is unlikely to be the major heating source during the late phase based on the available hard X-ray observations. In the long-lasting cooling scenario, the late phase evolution is more related to the total energy input, the density, and the loop length than heating mechanisms (thermal or non-thermal). In fact, we checked the effect of non-thermal beam heating using a flux of 10$^9$ erg cm$^{-2}$ s$^{-1}$ (the total energy is comparable with that of thermal heating), and found that the non-thermal effect is not significant in the EUV late phase. The increased density caused by a non-thermal beam could somewhat enhance the EUV late phase emission, but it affects the timing of the EUV late phase much less than a loop length variation. As for the beam heating term in the EBTEL model and its effects on plasma evolution, \cite{liuw13} gave a discussion in detail. The inputs of the EBTEL model are $Q(t)$ and $L$, and the outputs are $T(t)$ and $n(t)$. Using the outputs, we can calculate the synthetic EUV emissions and compare them with observations, such as the observed AIA fluxes and EVE intensities. In the EBTEL model, each loop is assumed to evolve independently with no interaction between different loops.

\subsection{Experiments in Three Cases}

We conduct three experiments based on observations. In cases 1 and 3, we consider two flare loops of quite different lengths to represent the main flaring loop and late phase loop in late phase flares. To check the validity of such a two-single-loop assumption, we also consider two sets of loops, in each of which the loop length and heating rate vary within a small amplitude, and obtain essentially similar results. Case 3 is different from case 1 by adding an additional heating to the late phase loop. In case 2, we consider a series of flare loops, rather than two distinct loops, with a main phase heating.

\subsubsection{Case 1: Main Phase Heating in Two Distinct Flare Loops}
\label{case1}

\begin{figure*}[htb]
\begin{center}
\includegraphics[width=16cm]{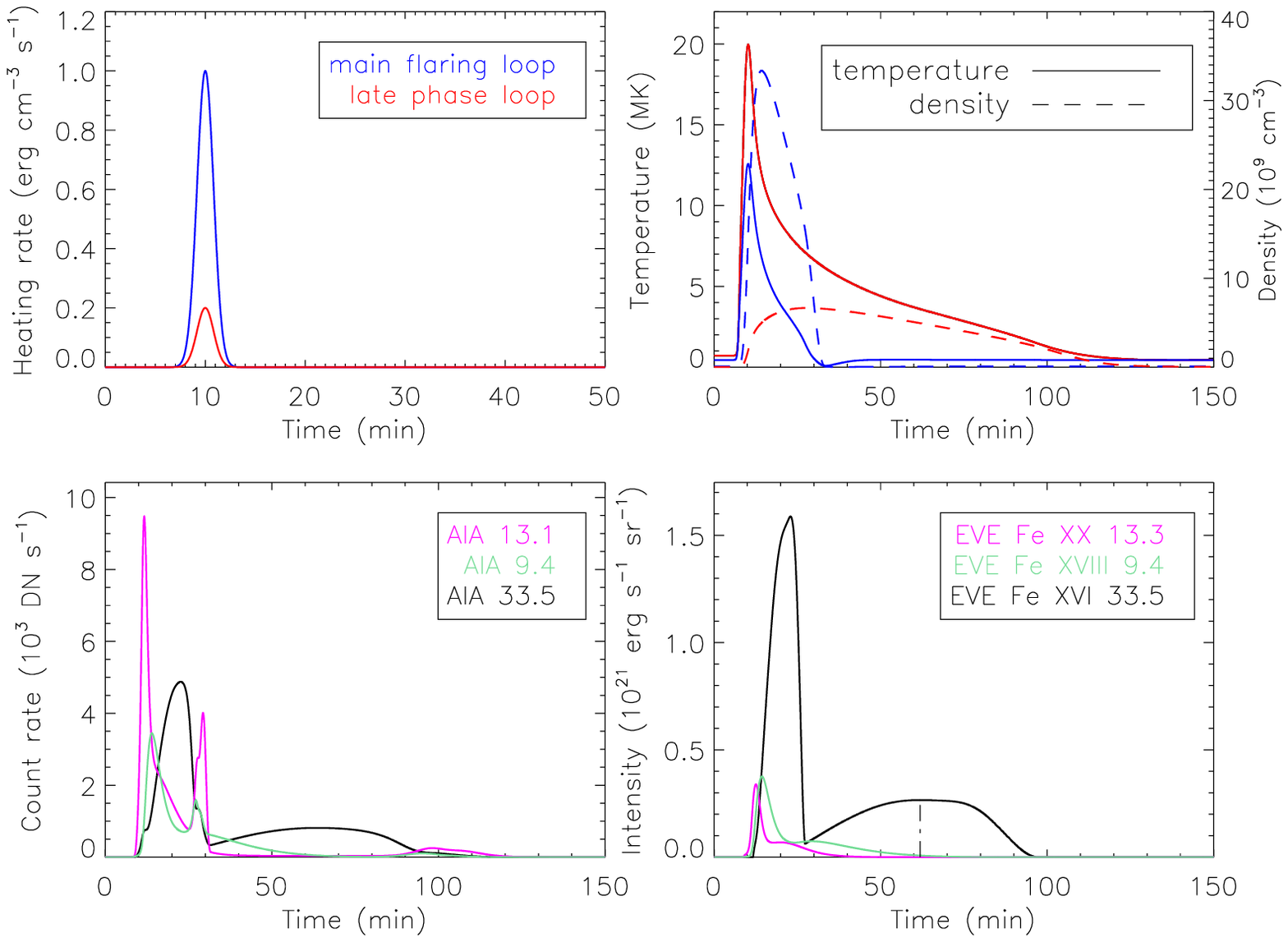}
\caption{Top left: heating profiles for two distinct flare loops. Top right: evolution of coronal mean temperature and density in these two loops. Bottom left: AIA fluxes in the three high-temperature channels by adding up emissions of the two loops. Bottom right: EVE intensities in three spectral lines, by adding up emissions of the two loops. The vertical dash-dot line marks the peak time of the EUV late phase.}
\label{evo1}
\end{center}
\end{figure*}

We first consider the case of main phase heating in two distinct flare loops, which have lengths markedly different (the length ratio is greater than 2). We consider that these two loops are magnetically connected, so that heating takes place in the two loops at the same time. The volumetric heating rates and loop half-lengths are listed in Table \ref{para1}. These values are not arbitrarily prescribed but chosen with the reference to observations or previous studies. Note that the loop length values are adopted from \cite{sunx13} and the heating rates are comparable with those given in \cite{qiuj12} and \cite{ying12,ying14}. Using the EBTEL model, we can compute the plasma evolution as well as the synthetic AIA filter fluxes and EVE line intensities. The results are shown in Table \ref{para1} and Figure \ref{evo1}.

\begin{table}[htb]
\begin{center}
\caption{Parameters of the Two Distinct Loops in Case 1}
\label{para1}
\begin{tabular}{cccccccc}
\tableline
\tableline
 Loops & $L$ & $Q_0$ & $t_0$ & $\tau$  & $Q_b$ & $T_{max}$ & $n_{max}$ \\ 
            & (Mm) & (erg cm$^{-3}$ s$^{-1}$) &   (min) & (s) & (erg cm$^{-3}$ s$^{-1}$) & (MK) & (10$^9$ cm$^{-3}$) \\
\hline
Main Flaring & 15 & 1.0 & 10 & 50 & 10$^{-5}$ & 13 & 33 \\
Late Phase & 75 & 0.2 & 10 & 50 & 10$^{-5}$ & 20 &  7 \\
\tableline
\end{tabular}
\end{center}
\small
$Q_0$ is the peak heating rate. The heating profile is Gaussian. The two loops have the same total heating energy.\\
$t_0$ is the timing of the peak heating rate.\\
$\tau$ is the Gaussian width, marking the heating timescale.\\
$Q_b$ is the background heating rate.\\
$T_{max}$ shows the peak mean temperature of the coronal loop.\\
$n_{max}$ shows the peak mean density of the coronal loop.
\end{table}

From Figure \ref{evo1} (top right panel), it is seen that the short main flaring loop evolves more quickly than the long late phase loop, especially in the decay (namely the cooling phase). We also compute the total AIA fluxes and EVE intensities from these two loops as a whole system (bottom panels). It can be seen that there are two peaks in the AIA and EVE light curves, particularly in the 33.5 nm emission. Note that there are some differences in AIA fluxes and corresponding EVE intensities, which are caused by the broad temperature response functions of the AIA channels. The secondary peak in the EVE Fe {\sc xvi} 33.5 emission occurs about 52 minutes after that in the AIA 13.1 nm emission (the latter usually peaks nearly at the same time as the {\em GOES} soft X-ray flux). Therefore, we have reproduced an EUV late phase in this experimental case, with heating only in the main phase in both the short (main flaring) and long (late phase) loops.

\begin{figure*}[htb]
\begin{center}
\includegraphics[width=16cm]{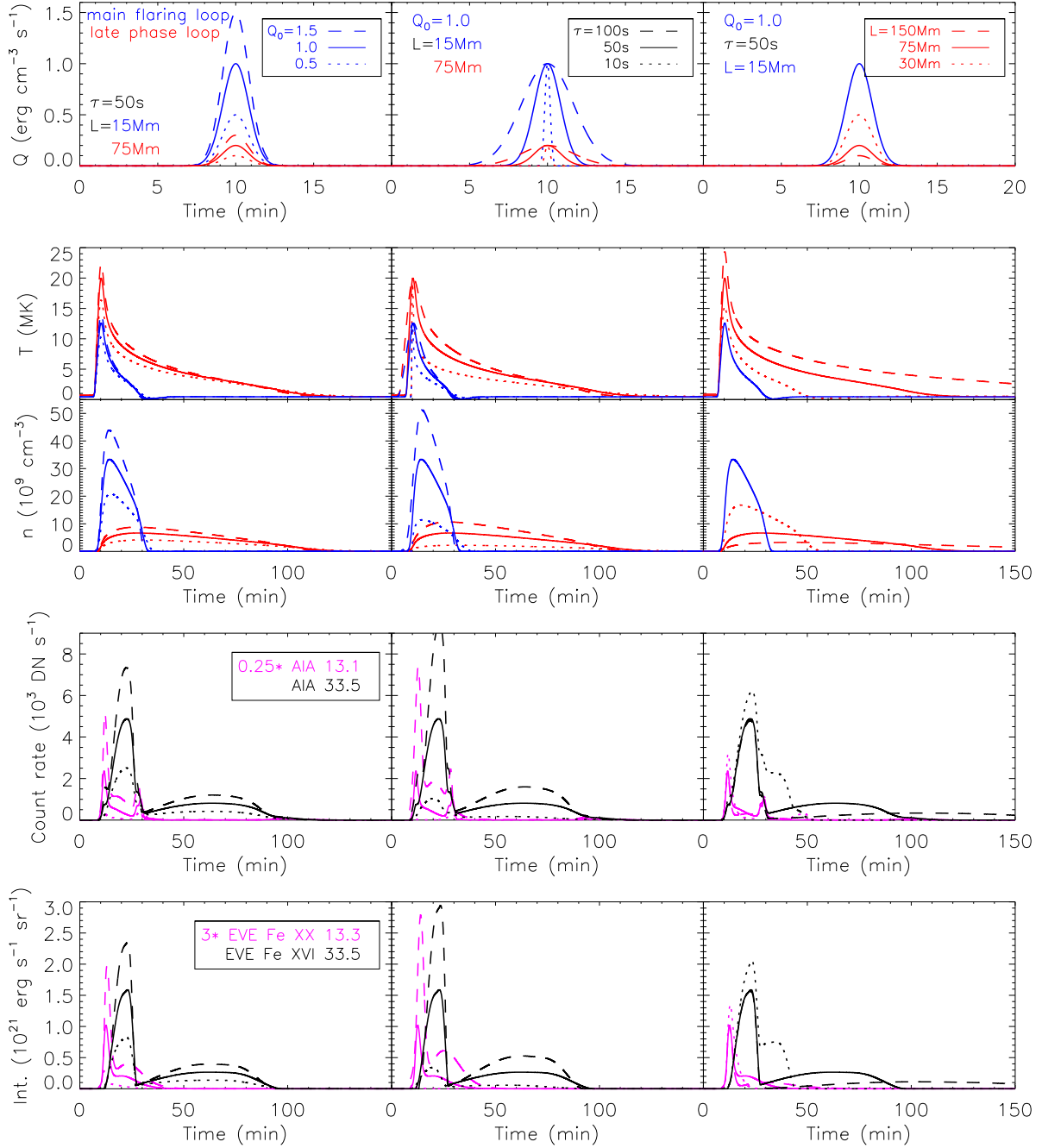}
\caption{Effects of varying loop parameters on plasma evolution and EUV emissions. Left: different peak heating rates ($Q_0$). Middle: different heating timescales ($\tau$). Right: different loop length ratios of late phase loop to main flaring loop. From top to bottom, plotted are the heating profiles for the main flaring and late phase loops, coronal mean temperature and density evolution of these two loops, synthetic AIA and EVE light curves in several bands by adding up emissions of these two loops.}
\label{para_survey}
\end{center}
\end{figure*}

We further check the effects of varying the input parameters (such as $Q_0$, $\tau$, and $L$) on plasma evolution and flare emission (in particular the late phase emission). Figure \ref{para_survey} shows the results of a parametric survey. We can see that, as the heating rate increases (the first column), the mean temperature and density of the plasmas in the main flaring and late phase loops increase as well; accordingly, the synthetic AIA fluxes and EVE intensities of the whole system get enhanced. It is interesting that, no matter how the heating rate changes, it only affects the magnitude of the late phase emission, but does not affect the timing of the late phase (as seen in particular for the synthetic EVE Fe {\sc xvi} 33.5 nm emission). On the other hand, if we increase the heating timescale, the temperature and density increase as well, although the peak temperature does not change much. Accordingly, the synthetic AIA fluxes and EVE intensities of the whole system are also enhanced. Similarly, varying the heating timescale mainly affects the magnitude of the late phase emission, but hardly affects its timing. However, when we change the length ratio of the late phase loop to the main flaring loop, the result is quite different. For example, we keep the length of main flaring loop unchanged ($L=$ 15 Mm) and vary the length of late phase loop to be 2, 5, and 10 times the former, respectively. Then we find that, as the length of late phase loop increases, the temperature increases, while the density decreases; and both of them evolve more slowly. More interesting, when the length ratio is relatively small (such as 2), the late phase is not well separated from the main phase. However, when the length ratio becomes large enough (such as 5 or 10), the late phase is well separated, although its peak is relatively weak. In the latter cases, the emission in the main phase is mainly from the main flaring loop, while the late phase emission comes from the late phase loop. It can be conjectured that as the length ratio becomes larger, the peak of the late phase appears later and weaker. In conclusion, the heating profile (including the heating rate and timescale) mainly affects the magnitude of the late phase, while the length of the late phase loop could affect both the magnitude and timing of the late phase.

\subsubsection{Case 2: Main Phase Heating in a Series of Flare Loops}
\label{2ndexp}

\begin{figure*}[htb]
\begin{center}
\includegraphics[width=16cm]{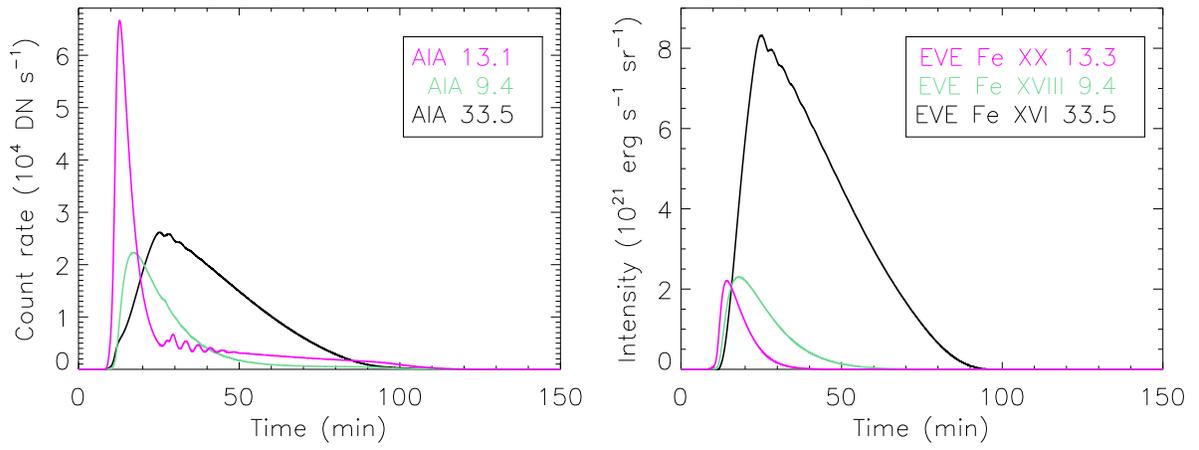}
\caption{Synthetic AIA fluxes (left) and EVE intensities (right) from 21 flare loops with main phase heating.}
\label{evo2}
\end{center}
\end{figure*}

Now we consider main phase heating in a series of flare loops, whose half-lengths increase from 15 Mm to 75 Mm with a step of 3 Mm. We have thus 21 flare loops in total rather than only two distinct loops. The parameters of the loops are listed in Table \ref{para2}. Inputting these parameters to the EBTEL model, we calculate the total emission of the whole flare loop system. Synthetic AIA and EVE light curves are shown in Figure \ref{evo2}. From the figure, it is seen that EUV late phase emission is not produced in this case. Therefore, we can conclude that the EUV late phase preferentially shows up in two sets of loops with a significant length difference between them rather than in a series of loops whose lengths are continuously varying when there only exists the main phase heating.

\begin{table}[htb]
\begin{center}
\caption{Parameters of the Loops in Case 2}
\label{para2}
\begin{tabular}{lccccc}
\tableline
\tableline
 Loops$^*$ & $L$ & $Q_0$ & $t_0$ & $\tau$ & $Q_b$  \\ 
 ($n$=21) & (Mm) & (erg cm$^{-3}$ s$^{-1}$) &  (min) & (s) & (erg cm$^{-3}$ s$^{-1}$) \\
\hline
 $i$=1 & 15 & 1.0 & 10 & 50 & 10$^{-5}$ \\
 $i$=2,...,20 & 15+3($i$-1) & 15/[15+3($i$-1)] & 10 & 50 & 10$^{-5}$ \\
 $i$=21 & 75 & 0.2 & 10 & 50 & 10$^{-5}$ \\
\tableline
\end{tabular}
\end{center}
\small
\ \ \ \ \ \ \ \ \ \ \ \ \ \ \ $^*$ $n$ is the number of the loops, and $i$ represents the $i$th loop.
\end{table}

\subsubsection{Case 3: Main and Late Phase Heatings in Two Distinct Flare Loops}

\begin{figure*}[htb]
\begin{center}
\includegraphics[width=16cm]{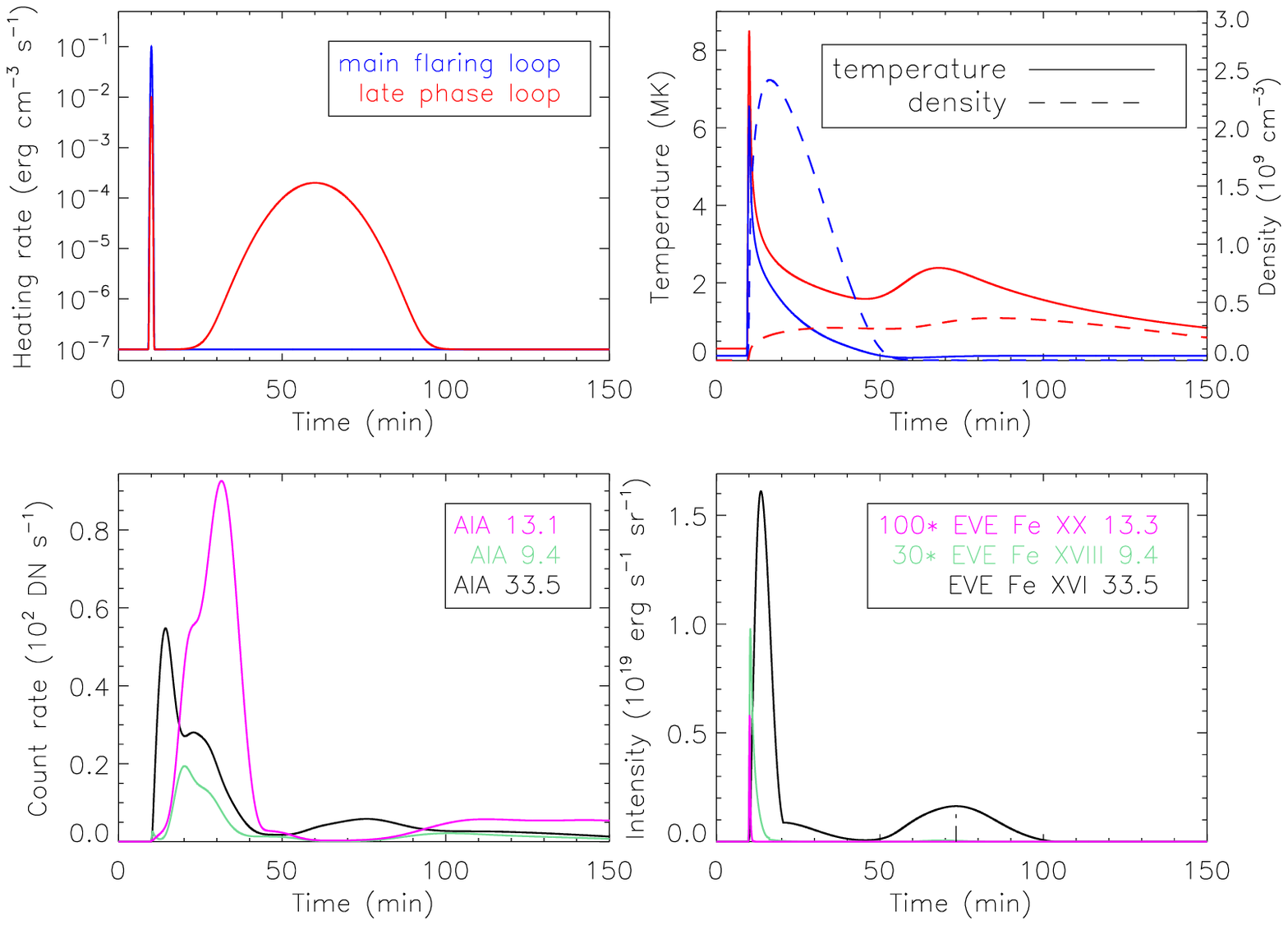}
\caption{Top left: heating profiles for two distinct flare loops. Top right: evolution of coronal mean temperature and density in these two loops. Bottom left: AIA fluxes in the three high-temperature channels by adding up emissions of the two loops. Bottom right: EVE intensities in three spectral lines, by adding up emissions of the two loops. The vertical dash-dot line marks the peak time of the EUV late phase.}
\label{evo3}
\end{center}
\end{figure*}

Finally, we consider main phase heating in both of the two loops plus a late phase (an additional) heating only in the late phase loop. The fraction of energy released in the late phase, if present, is a free parameter. In our simulations, we still assume that the total heating energy in the main flaring loop is equal to that in the late phase loop. The late phase loop is impulsively heated during the main phase, and then heated again and gradually during the decay phase, and the total heating energy in the main phase is the same as that in the decay phase. Note that the duration of the late phase heating is much longer than that of the main phase heating. The parameters of these two loops are listed in Table \ref{para3}. We plot the plasma evolution and synthetic emissions via EBTEL modeling in Figure \ref{evo3}. As seen in the figure, there is an enhancement in temperature (also in density) in the late phase loop tens of minutes after the main phase as a response to the long additional heating. Consequently, an EUV late phase shows up and peaks about one hour later than the main peak in the EVE 13.3 nm emission. Note that, in this case, the AIA 33.5 nm emission peaks first, followed by AIA 9.4 nm and then AIA 13.1 nm emission during the main phase, which is different from the usual cooling sequence of flare emissions. This may be caused by the very low heating energy in the main phase (the plasmas are not heated to 10 MK) combined with the broad temperature response functions of AIA filters. In this case, the peak time of the EUV late phase is related to the time of the additional heating. 

 \begin{table}[htb]
\begin{center}
\small
\caption{Parameters of the Two Distinct Loops in Case 3}
\label{para3}
\begin{tabular}{ccccccccc}
\tableline
\tableline
 Loops & $L$ & $Q_0$ & $t_0$ & $\tau$ & $Q_a$ & $t_a$ & $\tau_a$ & $Q_b$ \\ 
            & (Mm) & (erg cm$^{-3}$ s$^{-1}$)&  (min) & (s) & (erg cm$^{-3}$ s$^{-1}$) 
 & (min) & (s) & (erg cm$^{-3}$ s$^{-1}$) \\
\hline
Main Flaring & 15 & 0.1 & 10 & 10 & - & - & - & 10$^{-7}$ \\
Late Phase   & 75 & 0.01 & 10 & 10 & 0.0002 & 60 & 500 & 10$^{-7}$ \\
\tableline
\end{tabular}
\end{center}
\small
$Q_a$ is the peak heating rate of an additional heating in the late phase loop.\\
$t_a$ is the timing of the peak heating rate of additional heating.\\
$\tau_a$ is the Gaussian width, marking the heating timescale of additional heating.
\end{table}
 
In summary, the experiments show that, when heating only takes place in the main phase, an EUV late phase can be produced in two, or two sets of, distinct flare loops with quite different cooling timescales, but it is not produced in a series of flare loops with a continuous length variation. Continuous heating at a low rate in the decay phase also produces a late phase emission. We discuss the two scenarios of loop cooling and additional heating in the late phase in more detail in the following section.

\section{Nature of the EUV Late Phase}
\label{nature}

\subsection{Two Scenarios for Late Phase Emission}

Judging from our numerical experiments and observations, we see that the cooling process in flare loops is sufficient to explain the EUV late phase. In this scenario, there are two necessary conditions for the late phase: (1) the existence of two sets of magnetically related loop systems with distinctly different lengths and (2) sufficient plasma heating to temperatures higher than the warm coronal ones (e.g., 3 MK, the formation temperature of the EVE Fe {\sc xvi} 33.5 nm line). The first condition is one of the criteria for the EUV late phase in \cite{wood11}. In this case, the cooling timescales of the two sets of loops are quite different, resulting in an EUV late phase that can be well separated from the main phase. The second condition requires that the two sets of loops must both be heated to temperatures higher than 3 MK, the temperature at which the EUV late phase emission is most often observed; otherwise, one cannot see a secondary peak in the EVE Fe {\sc xvi} 33.5 nm emission.  

However, we cannot exclude the additional heating scenario. Additional heating in the decay phase may also produce the EUV late phase emission. In this case, the late phase appears when the additional heating comes into play. It may not require the late phase loop length to be much larger than the main flaring loop length. The situation may be different from case to case. In some events, such as the events analyzed by \cite{daiy13} and \cite{sunx13}, there does exist an additional heating that plays some role in producing the late phase. While in other events like that studied by \cite{liuk13}, it seems that no additional heating is required.

\subsubsection{Diagnostics of whether an Additional Heating Plays a Role or not in the EUV Late Phase}

\begin{figure*}[htb]
\begin{center}
\includegraphics[width=16cm]{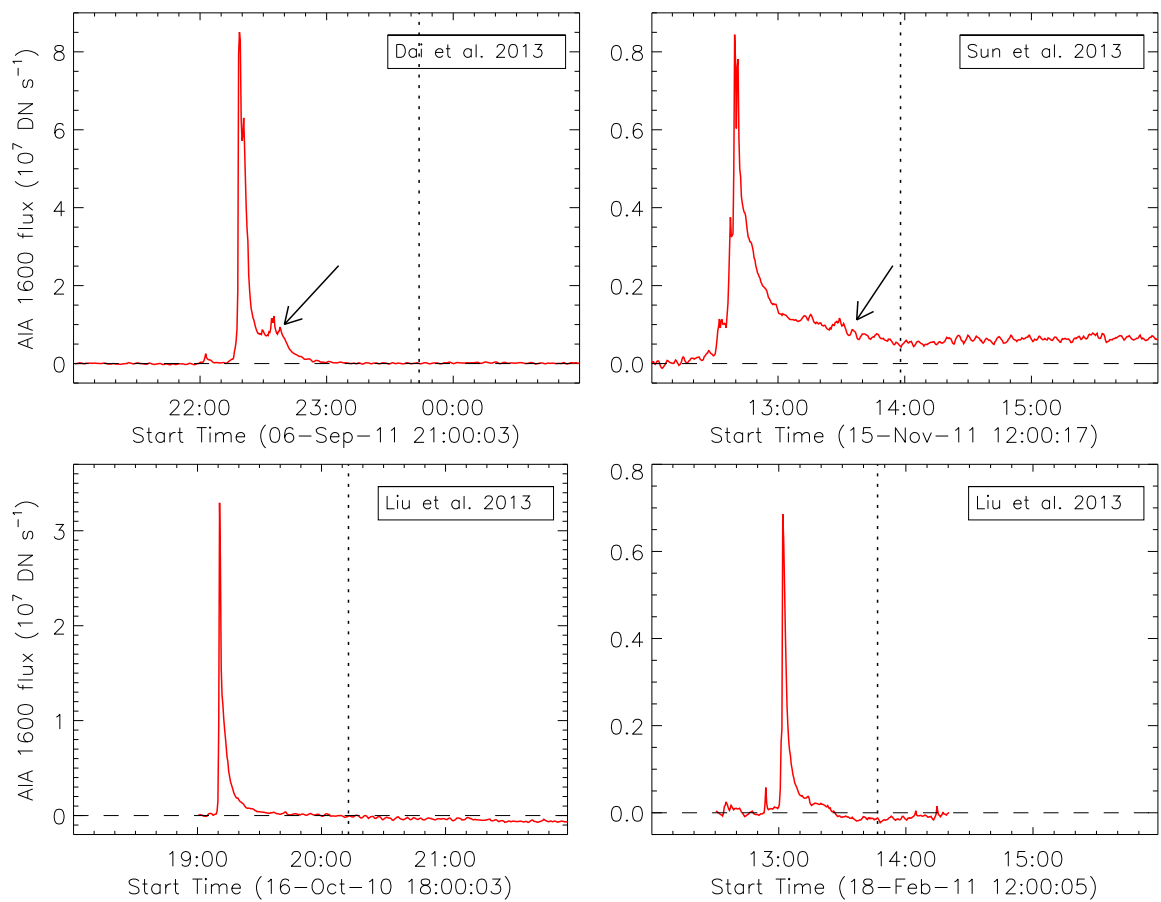}
\caption{AIA 1600 \AA~light curves over the whole AR for four late phase flare events. The horizontal dashed line in each panel marks the pre-flare emission level, and the vertical dotted line denotes the peak time of the EUV late phase in the EVE Fe {\sc xvi} 33.5 nm emission. The arrows show the additional heating signature. For clarity, all four panels are plotted with the same time interval (four hours).}
\label{AIA1600}
\end{center}
\end{figure*}

As discussed above, an additional heating could play a role in some late phase flares (e.g., \citealt{hock12,sunx13}), or not at all in other events (e.g., \citealt{liuk13}). How do we distinguish such different cases based on the observed emission features? Fortunately, the AIA UV 1600 \AA~emission provides a clue to judge the heating process. Emission in this channel includes the continuum formed in the temperature minimum region and the C {\sc iv} line formed in the upper chromosphere or transition region. During a flare, the optically-thin C {\sc iv} line is significantly enhanced. The AIA UV 1600 \AA~channel responds to the flare energy release very quickly and sensitively, and can be regarded as a good proxy of flare heating \citep{qiuj12, ying12, ying14}. We plot the light curves in AIA UV 1600 \AA~over the whole flaring AR for the four late phase flare events in \cite{daiy13}, \cite{sunx13}, and \cite{liuk13} in Figure \ref{AIA1600}. It is seen that, for the two events in which an additional heating plays a role (\citealt{daiy13,sunx13}), the UV light curves show an obvious hump later in the flare as marked by arrows; while for the events in \cite{liuk13} with no additional heating reported, there is no obvious enhancement in the decaying part of the short-duration light curves. Note that, the additional heating may occur somewhat before the late phase (e.g., \citealt{daiy13}), which might be called a delayed episode of heating; it can also last for quite a long time \citep{sunx13}. Therefore, the AIA UV 1600 \AA~emission can be used to determine whether additional heating plays a role or not in the EUV late phase emission.

There is another way to help diagnose the presence of additional heating. In the events of \cite{daiy13} and \cite{sunx13} with additional heating, the AIA EUV light curves of late phase loops show multiple peaks, especially in the low-temperature channels, e.g., 21.1 nm, 19.3 nm, and 17.1 nm, in the late phase (see Figure 4i in \citealt{daiy13} and Figure 7d in \citealt{sunx13}, respectively); while in the events of \cite{liuk13} without additional heating, the AIA EUV light curves of late phase loops show only one main peak during the late phase (see Figures 4 and 9 in \citealt{liuk13}). Note that, the multiple peaks are not obvious in AIA 33.5 nm emission because of the broad and flat temperature response function in this channel. The multiple peaks indicate multiple heating and cooling episodes and cannot be produced from the cooling of only the main phase heating.

The above two methods, based on the AIA UV 1600 \AA~and EUV emissions, can help to infer whether an additional heating plays a role or not in EUV late phase emission. In some cases, however, they are not sufficient for a definite conclusion. We expect to accumulate more late phase events in future to extract other diagnostic criteria that may help.

\subsection{Magnetic Configuration in the EUV Late Phase}

As discussed above, one of the conditions for the EUV late phase of flares is the existence of two sets of magnetically related loop systems. This means that EUV late phase flares have multipolar magnetic configurations, which have been mentioned in previous studies. In particular, \cite{sunx13} proposed a spine-fan topology for the M1.9 flare on 15 November 2011, and the late phase is assumed to be caused by the hot spine structure. However, some of the magnetic configurations are only inferred from coronal loop observations, e.g., the M2.9 and M1.4 flares in \cite{liuk13} and the X2.1 flare in \cite{daiy13}, but not studied in detail with 3D magnetic field models. Here, we provide the 3D magnetic topology for the four late phase flares (the M2.9, M1.4, X2.1, and M1.9 flares) with a NLFFF model. Although \cite{sunx13} computed the magnetic topology for the M1.9 flare, we show it here again for comparison.

We adopt the optimization method proposed by \citet{2000Wheatland} and implemented by \citet{2004Wiegelmann} to perform NLFFF modeling. The optimization method uses a vector magnetic field as the bottom boundary condition, which is provided by the Helioseismic and Magnetic Imager \citep[HMI;][]{2012Scherrer,2012Schou} on board \textit{SDO}. The $180^\circ$ ambiguity in the transverse components of the vector magnetic field is resolved by an improved version of the minimum energy method \citep{2006Metcalf,2009Leka}. Then, we correct the projection effect following the method proposed by \citet{1990Gary}; therefore, the magnetic field vectors are projected onto vertical and horizontal components and the geometry is mapped onto a surface tangent to the center of a selected field of view. The NLFFF model also requires that the net magnetic force and torque in an isolated magnetic domain on the boundary vanish. The vector magnetic field is preprocessed with the method of \citet{2006Wiegelmann} to satisfy the aforementioned conditions. Finally, we use the preprocessed vector magnetic field as the bottom boundary for the NLFFF modeling.

\begin{figure*}[htb]
\begin{center}
\includegraphics[width=13cm]{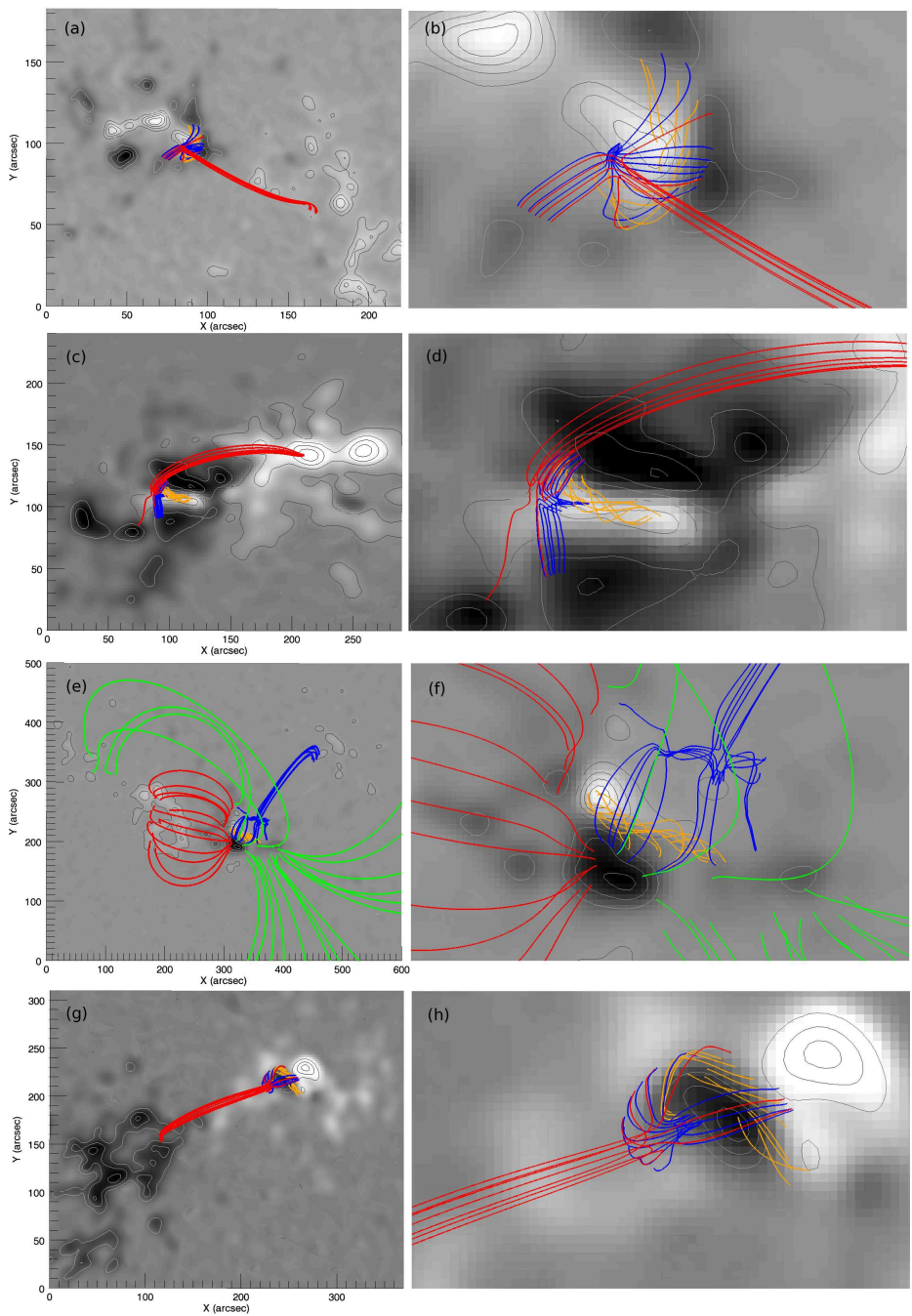}
\caption{Magnetic field topology of the M2.9, M1.4, X2.1, and M1.9 flares at 16 October 2010 18:36 UT, 18 February 2011 12:24 UT, 6 September 2011 21:48 UT, and 15 November 2011 12:12 UT from top to bottom row, respectively. Grey-scale images and contours mark the vertical magnetic field on the bottom boundary. Orange, blue, red, and green solid lines indicate magnetic field lines of the NLFFF model. In particular, red lines indicate those magnetic field lines contributing to the late phase of the flare. The left column shows a larger field of view, while the right one shows a zoomed-in field of view.}
\label{mag}
\end{center}
\end{figure*}

The magnetic field topologies of the four flares are plotted in Figure \ref{mag}. Comparing the vertical magnetic field in the source region of the four flares, we find a common feature, namely, the presence of one polarity that is fully surrounded by the opposite polarity. For example, the positive polarity in the source region for the M2.9, M1.4, and X2.1 flares is surrounded by negative polarity (first three rows of Figure \ref{mag}), and the negative polarity is surrounded by the positive one for the M1.9 flare (last row of Figure \ref{mag}). This is strong evidence for the existence of a magnetic null point in the corona. We calculate the positions of the null points quantitatively by solving the equation $\mathbf{B}(x,y,z) = 0$ with a modified Powell hybrid method \footnote{http://www.lesia.obspm.fr/fromage/}. The linear structure, namely the fan and spine, of the null point is determined by the eigenvalues and eigenvectors of the Jacobian matrix of the magnetic field in the vicinity of the null point. The spine is an isolated field line that links to the null point, while the the fan is a surface of field lines that link to the null and spread away from it as a separatrix surface \citep{prie96}. We plot the magnetic field lines around the null point with blue and red lines for the M2.9, M1.4, M1.9 flares, and only with blue lines for the X2.1 flare in Figure \ref{mag}.

Comparing the EUV observations in \cite{liuk13} with the magnetic topologies in the first two rows, one can find that both the late phases for the M2.9 and M1.4 flares are contributed by the hot spines. This conclusion is similar to that in \cite{sunx13}, but was not pointed out in \cite{liuk13}. Note that, there is an interesting difference between the events in \cite{daiy13} and \cite{sunx13}. Although there existed a magnetic null point for the event in \cite{daiy13}, the late phase emission did not originate from the spine-fan loops but from the large loops in the eastern side as shown by the red magnetic field lines in the third row of Figure \ref{mag}. This point has been proven in detail with the spatially resolved EUV observations in \cite{daiy13}. In conclusion, two sets of magnetically related loops with quite different loop lengths can be generated by either spine field lines associated with a magnetic null point or large-scale magnetic loops in multipolar magnetic fields. Both magnetic configurations could produce an EUV late phase flare.

\subsection{Why does the EUV Late Phase Appear in Warm Coronal Emissions} 

The EUV late phase of flares is usually observed in warm coronal emissions, such as the EVE Fe {\sc xvi} 33.5 nm and Fe {\sc xv} 28.4 nm lines, while it is not clearly seen in hot emissions, like the {\em GOES} soft X-ray and the EVE Fe {\sc xx} 13.3 nm line. \cite{liuk13} proposed that, on one hand, the hot emissions of the EUV late phase are much smaller than the impulsive (or main) phase; on the other hand, the EUV late phase overlaps with the main phase in time. Our numerical experiments basically support those explanations. According to the theory of chromospheric evaporation, the emissions in EUV bands originate from hot evaporated plasmas that cool down gradually until they are not emissive in a specific wavelength. In the initial cooling process, the conductive cooling dominates, with a rate proportional to loop length squared \citep{carg95,sunx13}. Therefore, the cooling time differs for the two sets of loops since they have different lengths. The shorter main flaring loops cool faster and the longer late phase loops cool slower. If the time difference for plasma to cool down to a temperature responsible for warm emissions, especially the EVE Fe {\sc xvi} 33.5 nm, is large enough, the late phase emission can be well separated from the main phase emission, resulting in a late phase signature in these emissions. In the hot emissions, however, the time difference for cooling to the corresponding temperature is too short to separate the contributions from the two sets of loops in convolved light curves. There is another possibility that the late phase loops are only heated to the warm temperatures; then one cannot observe the late phase emission in hot channels. 

In fact, although the EUV late phase of flares appears most preferentially in the EVE Fe {\sc xvi} 33.5 nm emission, one can also detect a late phase signature in some cool EVE lines, such as Fe {\sc xi} 18.0 nm and Fe {\sc x} 17.7 nm ($\sim$1 MK), at least in some flares (e.g., \citealt{wood11}). Usually, the late phase peak in cool emissions appears later than that in warm emissions, which is consistent with the loop cooling scenario. In some other late phase flares, however, one cannot see an obvious late phase emission in cool EVE lines (e.g., \citealt{daiy13}). In this case, it is probable that the cool EUV emission in the late phase is rather weak and submerged in the bulk coronal emission \citep{liuk13} or that there appear coronal dimmings in these lines due to mass depletion caused by CME lift-off. 

From the disk-integrated observations of EVE, one may not be able to detect the late phase emission in all the EUV lines; one can observe the late phase arcades in nearly all the AIA EUV channels (e.g., the hot 13.1 nm and 9.4 nm bands, the warm 33.5 nm band, and the cool 21.1 nm, 19.3 nm, and 17.1 nm bands) in spatially resolved images. In addition, the broad temperature response functions of AIA also make it favorable to observe the bright late phase loops in EUV images.

\subsection{Why do not All Flares Exhibit an EUV Late Phase}

Finally, one may naturally ask a question why the EUV late phase appears only in a fraction of solar flares, but not in others. This question is challenging since loop cooling occurs in all cases. We propose two possible explanations here. One possibility is that flare heating takes place in loops with a continuous length distribution, as discussed above. The cooling time of these loops also changes smoothly. Therefore, the emissions from the loops overlap in time and cannot be clearly separated from each other; observationally, one cannot detect a second emission peak (see the second experiment in Section \ref{2ndexp}). Many solar flares should fit this case. In particular, if there is a continuous heating in a series of loops, it may correspond to a long duration event. The second possibility is mainly for the flares that have two sets of loops with quite different lengths. If these two sets of loops are not heated to warm coronal temperatures (e.g., 3 MK, the formation temperature of the EVE Fe {\sc xvi} 33.5 nm line), one is naturally unable to observe an EUV late phase in warm coronal emissions either.

\section{Summary and Discussions}
\label{discussion}

We have investigated the EUV late phase of solar flares with the help of numerical experiments and {\em SDO} observations. Using the EBTEL model, we study the role of long-lasting cooling and additional heating in the appearance of an EUV late phase. Our results show that the long cooling process in late phase loops is a preferable mechanism responsible for the flare late phase. An additional heating , however, may exist in some cases. To distinguish whether the additional heating plays a role or not in the EUV late phase, we provide some preliminary methods based on AIA UV and EUV emission features. In the case of the long-lasting cooling scenario, there are two necessary conditions for an EUV late phase: one is the existence of two sets of magnetically related loop systems of different lengths, and the other is a strong enough heating for both sets of loops to reach warm coronal temperatures. We also explain the reason why the late phase shows up most obviously in warm coronal emissions (e.g., the Fe {\sc xvi} 33.5 nm line): it is a result of different cooling timescales for the main and late phase loops. Furthermore, we have used NLFFF modeling to propose that the EUV late phase can be generated either in hot spine field lines associated with a magnetic null point or in large-scale magnetic loops in multipolar magnetic fields.

Although the 0D EBTEL model is based on simple assumptions, it is appropriate to use the model to study the EUV late phase in flares. The primary limitation is that the model only computes mean quantities of  temperature and density but not their distributions along the loop. This mean property, however, mainly affects the slopes of synthetic AIA and EVE light curves, while it has little effects on the appearance of an EUV late phase emission. It has been shown that the simple EBTEL model can indeed capture the main characteristics of evolution of flare plasmas \citep{klim08,carg12,qiuj12,ying12,ying14,liuw13}.

To reveal the nature of EUV late phase emission helps us to better understand the energy process of the whole flare. In fact, whether there is a new energy release process (magnetic reconnection, particle acceleration etc.) or not in the late phase is of interest for space weather forecast.


\acknowledgments
The authors are very grateful to the anonymous referee for many valuable comments and to Eric Priest for a thorough reading of the manuscript. Y.L. thanks Jiong Qiu and Tongjiang Wang for their scientific discussions. This project was supported by NSFC under grants 10933003, 11373023, 11203014, 11103009, and 11103008 and by NKBRSF under grants 2011CB811402 and 2014CB744203. Y.L. is also supported by Jiangsu Province under grant 0201003404. \textit{SDO} is a mission of NASA's Living With a Star Program.

\bibliographystyle{apj}

\end{document}